\documentclass[aps,prb,twocolumn,epsfig,showpacs,preprintnumbers,superscriptaddress,float]{revtex4}
\usepackage{graphics}
\usepackage{graphicx}
\usepackage{dcolumn}
\usepackage{bm}
\usepackage{amssymb}

\usepackage{amsmath}

\begin{document}
\title
{Large anomalous Nernst and spin Nernst effects in noncollinear antiferromagnets Mn$_3X$ ($X$ = Sn, Ge, Ga)}
 
\author{Guang-Yu Guo}
\email{gyguo@phys.ntu.edu.tw}
\affiliation{Department of Physics and Center for Theoretical Physics, National Taiwan University, Taipei 10617, Taiwan}
\affiliation{Physics Division, National Center for Theoretical Sciences, Hsinchu 30013, Taiwan}
\author{Tzu-Cheng Wang}
\affiliation{Department of Physics and Center for Theoretical Sciences, National Taiwan University, Taipei 10617, Taiwan}

\date{\today}   

\begin{abstract}
Noncollinear antiferromagnets have recently been attracting considerable interest partly
due to recent surprising discoveries of the anomalous Hall effect (AHE) in them and partly
because they have promising applications in antiferromagnetic spintronics. 
Here we study the anomalous Nernst effect (ANE), a phenomenon having the same origin as
the AHE, and also the spin Nernst effect (SNE) as well as AHE and the spin Hall effect (SHE)
in noncollinear antiferromagnetic Mn$_3X$ ($X$ = Sn, Ge, Ga) within the Berry 
phase formalism based on {\it ab initio} relativistic band structure calculations. 
For comparison, we also calculate the anomalous Nernst conductivity (ANC) and anomalous Hall
conductivity (AHC) of ferromagnetic iron as well as the spin Nernst conductivity (SNC) of
platinum metal. Remarkably, the calculated ANC at room temperature (300 K)
for all three alloys is large, being up to 5 times larger than that of iron. 
Moreover, the calculated SNC for Mn$_3$Sn and Mn$_3$Ga is also large, being as large 
as that of platinum. This suggests that these antiferromagnets would
be useful materials for thermoelectronic devices and spin caloritronic devices.
The calculated ANC of Mn$_3$Sn and iron are in reasonably good agreement with
the very recent experiments. The calculated SNC of platinum also agrees with
the very recent experiments in both sign and magnitude.
The calculated thermoelectric and thermomagnetic properties are analyzed 
in terms of the band structures as well as the energy-dependent AHC, ANC, SNC and spin Hall 
conductivity via the Mott relations. 
\end{abstract}
 
\maketitle

\section{Introduction}
In recent two decades, spin transport electronics (spintronics) has attracted enormous interest 
because of its promising applications in information storage and processing and other electronic 
technologies\cite{Prin98,Zuti04}. Spin current generation, detection and manipulation are 
three key issues in the spintronics. In this context, spin-related transport phenomena in 
solids especially in those materials that can provide highly spin-polarized
charge current and large pure spin current, have been intensively investigated recently.
The anomalous Hall effect (AHE), discovered in 1881 by Hall \cite{Hall81}, and the 
spin Hall effect (SHE), predicted in 1971 by Dyakonov and Perel \cite{Dyak71}, are two principal 
spin-related transports and thus have received renewed interests.\cite{Naga10,Sino15}  
Intuitively, spin-up and spin-down electrons moving through the relativistic band structure of a solid 
experience opposite transverse velocities caused by an applied electric field.
In a ferromagnet where an unbalance of spin-up and spin-down electrons exists, these opposite currents result in
a spin-polarized transverse charge current and hence the (intrinsic) AHE. Therefore, the AHE is usually
assumed to be proportional to the magnetization of the magnetic material. In a nonmagnetic material where 
spin-up and spin-down electrons are equal in numbers, this process gives rise to a pure transverse spin current, and 
this is known as the (intrinsic) SHE. The SHE is particularly important for spintronics because it enables us
to generate, detect and control spin current without magnetic field or magnetic materials.\cite{Hoff13,Sino15} 
Furthermore, the pure spin current is dissipationless\cite{Mura03} and is thus especially useful 
for the development of low power-consumption nanoscale spintronic devices\cite{Liu12}.

Interestingly, Chen {\it et al.} \cite{Chen14} recently showed that large AHE could occur
in noncollinear antiferromagnets without net magnetization such as cubic Mn$_3$Ir. This surprising 
result arises from the fact that in a three-sublattice kagome lattice with a noncollinear
triangle antiferromagnetic structure, not only the time-reversal symmetry ($\mathcal{T}$)
is broken but also there is no spatial symmetry operation ($\mathcal{S}$) which, in conjunction
with $\mathcal{T}$, is a good symmetry that preserves the Kramers theorem.
Subsequently, large AHE was observed in hexagonal noncollinear antiferromagnets
Mn$_3$Sn\cite{Nak15}  and Mn$_3$Ge\cite{Kiy16,Nay16}. In the meantime, large SHE was
predicted in noncolinear antiferromagnets Mn$_3X$ ($X =$ Sn, Ge, Ir)\cite{Zha16,Zha17}
and was also observed in Mn$_3$Ir\cite{Zha16}. All these fascinating findings suggest that
these noncollinear antiferromagnets may find exciting applications in spintronics,
an emergent field called antiferromagnetic spintronics\cite{Jung16}. Antiferromagnetic spintronics
has been attracting increasing attention in recent years because antiferromagnetic materials
have several advantages over ferromagnetic materials. In particular, antiferromagnetic elements
would not magnetically affact their neighbors and are insensitive to stray magnetic fields. 
Moreover, antiferromagnets have faster spin dynamics than ferromagnets, and this would lead to ultrafast
data processing.

In a ferromagnet, the charge Hall current could also arise when a temperature gradient ($\nabla T$) 
instead of an electric field, is applied. This phenomenon, due to the simultaneous presence 
of the spin-orbit coupling (SOC) and net magnetization in the ferromagnet, is refered to as the anomalous Nernst effect 
(ANE)\cite{Nern87,Lee04,Xiao06}. Similarly, a temperature gradient could also generate the 
spin Hall current in a nonmagnetic material, and this is known as the spin Nernst effect 
(SNE)\cite{Chen08}. Clearly, materials that exhibit large ANE and SNE would have useful applications for
spin thermoelectronic devices driven by heat, a new field known as spin caloritronics\cite{Bau12}.
This offers exciting prospects of 'green' spintronics powered by, e.g., waste ohmic heat.
Since the ANE and SNE, respectively, have the same physical origins as the AHE and SHE, one 
could expect significant ANE and SNE in the above-mentioned noncollinear antiferromagents as well.
In other words, noncollinear antiferromagnets Mn$_3X$ ($X =$ Sn, Ge, Ga) could also be
useful materials for developing antiferromagnetic spin caloritronics. 
Nevertheless, no investigation of the SNE in noncollinear antiferromagnets has been reported
and only two reports on the measurement of the ANE in Mn$_3$Sn appeared very recently.\cite{Li16,Ikhl17}

In this paper, therefore, we perform an {\it ab initio} study on the ANE and SNE
in hexagonal noncollinear antiferromagnets Mn$_{3}$Ga, Mn$_{3}$Ge and Mn$_{3}$Sn (Fig. 1),
based on the density functional theory (DFT) with the generalized gradient approximation 
(GGA)\cite{Perdew1996}. For comparison, we also study the ANE in bcc Fe, a ferromagnetic 
transition metal having large AHE\cite{Yao04}, and the SNE in fcc Pt, a heavy nonmagnetic 
transition metal exhibiting gigantic SHE\cite{Guo08}. Indeed,
we find that the anomalous Nernst conductivity at room temperature of all three alloys
is large, being up to five times larger than that of bcc Fe.
The spin Nernst conductivity of Mn$_{3}$Ga and Mn$_{3}$Sn is as large as that of fcc Pt. 
The rest of this paper is organized as follows.
In the next section, we briefly describe the Berry phase formalism for 
calculating the intrinsic Hall and Nernst conductivities as well as the computational details.
Section III consists of three subsections. 
We first present the calculated total energy and magnetic 
properties of two low-energy noncollinear antiferromagnetic structures 
[Figs. 1(c) and 1(d)] of Mn$_{3}X$ and also compared our results with
available previous experimental and theoretical reports in Subsec. III A. 
We then report the calculated anomalous Nernst
conductivity as well as anomalous Hall conductivity for these magnetic structures in Subsec. III B.
We finally present the calculated spin Nernst conductivity and also spin Hall conductivity
in Subsec. III C. Finally, the conclusions drawn from this work are summarized in Sec. IV.

\section{Theory and Computational Method}
Here we consider ordered Mn$_{3}$Ga, Mn$_{3}$Ge and Mn$_{3}$Sn alloys in 
the layered hexagonal DO$_{19}$ ($P6_3/mmc$ or $D_{6h}^4$) structure [see Fig. 1(a)] and
use the experimental lattice constants of $a = 5.36$ \AA$ $ and $c = 4.33$ \AA$ $~\cite{Kre70}, 
$a = 5.34$ \AA$ $ and $c = 4.31$ \AA$ $~\cite{Kiy16}, and $a = 5.66$ \AA$ $ 
and $c = 4.53$ \AA$ $~\cite{Nak15}, respectively.
The primitive unit cell contains two layers of Mn triangles stacked along the $c$-axis,
and in each layer the three Mn atoms form a kagome lattice with the $X$ atom located 
at the center of each hexagon (Fig. 1). 
The total energy and electronic structure are calculated based on the DFT
with the GGA in the form of Perdew-Burke-Ernzerhof~\cite{Perdew1996}.
The accurate projector-augmented wave (PAW) method~\cite{Blochl1994},
as implemented in the Vienna \textit{ab initio} simulation 
package (\textsc{vasp})~\cite{Kresse1993,Kresse1996}, is used.
The fully relativistic PAW potentials are adopted in order to include the SOC.
The valence configurations of Mn, Sn, Ge and Ga atoms taken into account in the calculations
are $3d^{6}4s^{1}$, $4d^{10}5s^{2}5p^2$, $3d^{10}4s^{2}4p^2$ and $3d^{10}4s^{2}4p^1$, respectively.
A large plane-wave cutoff energy of 350 eV is used throughout. 
In the self-consistent electronic structure calculations, a fine $\Gamma$-centered $\bm{k}$-point mesh
of 20$\times$20$\times$20 [i.e., 2112 $k$-points over the orthorhombic irreducible Brillouin 
zone wedge (IBZW) (see Fig. 1)] is adopted for the Brillouin zone (BZ) integration using the
tetrahedron method\cite{Jeps71}.

\begin{figure}
\includegraphics[width=\columnwidth]{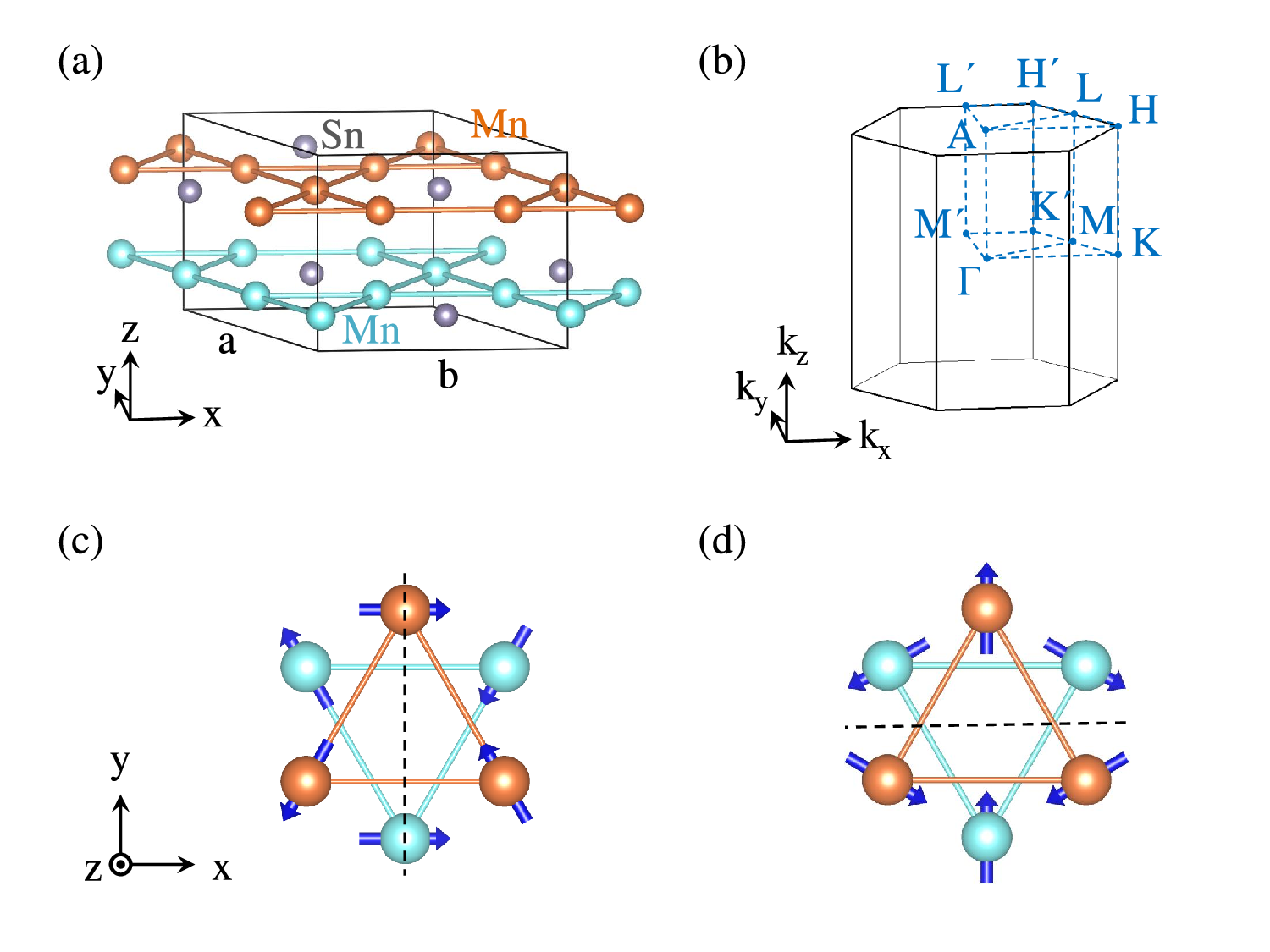}
\caption{(Color online) (a) Layered hexagonal ($D_{6h}^4$) structure of Mn$_{3}X$ ($X$ = Sn, Ge, Ga)
with (b) the associated hexagonal Brillouin zone (BZ). (c) Type A and (d) type B antiferromagnetic 
configurations considered in this paper. Both magnetic structures have an orthorhombic symmetry
and thus their irreducible BZ wedge (IBZW) [i.e., the trapzian prism indicated by the blue
dashed lines in (b)] is three times larger than the hexagonal IBZW [i.e., the triangle prism 
indicated by the blue dashed lines in (b)]. The vertical [horizontal] black dashed line in (c) [(d)]
denotes the mirror plane. } 
\label{fig:crystal}
\end{figure}

The anomalous Hall conductivity (AHC) and anomalous Nernst conductivity (ANC) are calculated
based on the elegant Berry-phase formalism\cite{Xiao10}. Within this Berry-phase formalism, 
the AHC ($\sigma_{ij}^{A} = J^c_i/E_j$) is 
simply given as a BZ integration of the Berry curvature for all the occupied bands,
\begin{eqnarray}
\sigma_{ij}^{A} = -\frac{e^2}{\hbar}\sum_n \int_{BZ}\frac{d{\bf k}}{(2\pi)^3}f_{{\bf k}n}\Omega_{ij}^n({\bf k}),\nonumber \\
\Omega_{ij}^n({\bf k}) = -\sum_{n'\neq n}
\frac{2{\rm Im}[\langle{\bf k}n|v_i|{\bf k}n'\rangle\langle{\bf k}n'|v_j|{\bf k}n\rangle]}
 {(\epsilon_{{\bf k}n}-\epsilon_{{\bf k}n'})^2},
\end{eqnarray}
where $f_{{\bf k}n}$ and ${\Omega_{ij}^n({\bf k})}$ are the Fermi distribution function and the Berry curvature 
for the $n$th band at ${\bf k}$, respectively. $i$ and $j$ $\in (x,y,z)$, and $i \neq  j$.  
$J^c_i$ is the $i$-component of the charge current density ${\bf J}^c$ and $E_j$ is the $j$-component of
the electric field ${\bf E}$.
Moreover, the ANC ($\alpha_{ij}^{A} = -J^c_i/\nabla_jT$) can be written as
\begin{eqnarray}
\alpha_{ij}^A = \frac{1}{T}\frac{e}{\hbar}\sum_n \int_{BZ}\frac{d{\bf k}}{(2\pi)^3}\Omega_{ij}^n({\bf k})\nonumber \\
\times [(\epsilon_{{\bf k}n}-\mu)f_{{\bf k}n}+k_BT\textrm{ln}(1+e^{-\beta(\epsilon_{{\bf k}n}-\mu)})],
\end{eqnarray} 
where $\mu$ is the chemical potential and $k_B$ is the Boltzmann constant.\cite{Xiao06,Guo14}

The Berry curvature ${\Omega_{ij}^n({\bf k})}$ can be considered as a pseudovector\cite{Xiao10}, 
just like the spin, and thus can be written as 
${[\Omega_n^x({\bf k}),\Omega_n^y({\bf k}),\Omega_n^z({\bf k})] = [\Omega_{yz}^n({\bf k}),\Omega_{zx}^n({\bf k}),\Omega_{xy}^n({\bf k})]}$. 
Thus, ${{\bf \Omega}_n({\bf k}) = {\bf \Omega}_n(-{\bf k})}$ if the system has spatial 
inversion symmetry ($\mathcal{P}$) and ${{\bf \Omega}_n({\bf k}) = -{\bf \Omega}_n(-{\bf k})}$ 
if it has $\mathcal{T}$ symmetry.\cite{Xiao10}
Obviously, if the system has both $\mathcal{P}$ and $\mathcal{T}$ symmetries,
${{\bf \Omega}_n({\bf k})}$ becomes identically zero. The AHC and ANC are also pseudovectors and can be written
as ${[\sigma_A^x,\sigma_A^y,\sigma_A^z] = [\sigma_{yz}^A,\sigma_{zx}^A,\sigma_{xy}^A]}$ and
${[\alpha_A^x,\alpha_A^y,\alpha_A^z] = [\alpha_{yz}^A,\alpha_{zx}^A,\alpha_{xy}^A]}$, respectively.

Similarly, the spin Hall conductivity ($\sigma_{ij}^{s} = J^s_i/E_j$) is given by a BZ integration of the 
spin Berry curvature ($\Omega_{ij}^{n,s}({\bf k})$) for all the occupied bands,
\begin{eqnarray}
\sigma_{ij}^{s} = -e\sum_n \int_{BZ}\frac{d{\bf k}}{(2\pi)^3}f_{{\bf k}n}\Omega_{ij}^{n,s}({\bf k}),\nonumber \\
\Omega_{ij}^{n,s}({\bf k}) = -\sum_{n'\neq n}
\frac{2{\rm Im}[\langle{\bf k}n|\{\tau_s,v_i\}/4|{\bf k}n'\rangle\langle{\bf k}n'|v_j|{\bf k}n\rangle]}
 {(\epsilon_{{\bf k}n}-\epsilon_{{\bf k}n'})^2},
\end{eqnarray}
where $s$ denotes the spin direction and $\tau_s$ is a Pauli matrix.\cite{Guo08}
Then the spin Nernst conductivity ($\alpha_{ij}^{s} = -J^s_i/\nabla_jT$) can be written as
\begin{eqnarray}
\alpha_{ij}^s = \frac{1}{T} \sum_n \int_{BZ}\frac{d{\bf k}}{(2\pi)^3}\Omega_{ij}^s({\bf k}n)\nonumber \\
\times [(\epsilon_{{\bf k}n}-\mu)f_{{\bf k}n}+k_BT\textrm{ln}(1+e^{-\beta(\epsilon_{{\bf k}n}-\mu)})],
\end{eqnarray} 
where $J^s_i$ denotes the $i$-component of the spin current density ${\bf j}^s$ with spin being along the $s$-axis.

In the AHC, SHC, ANC and SNC calculations, the velocity ($\langle{\bf k}n|v_i|{\bf k}n'\rangle$) 
and spin-velocity ($\langle{\bf k}n|\{\tau_s,v_i\}/4|{\bf k}n'\rangle$) matrix elements 
are obtained from the self-consistent electronic structure within the PAW formalism.\cite{Adol01} 
To obtain accurate AHC, SHC, ANC and SNC, a dense $k$-point mesh would be needed.\cite{Yao04,Guo05} 
Therefore, we use a very fine mesh of 97344 k-points on the magnetic IBZW (1/8 BZ), together with the tetrahedron
method\cite{Jeps71}. This is equivalent to a large number of $k$-points of $\sim$778752 in the full BZ, 
and corresponds to the division of the $\Gamma$K line into $n_d$ = 50 intervals. Further
calculations using $n_d$ = 20, 30 and 40 (i.e., 7260, 22272, 51597 $k$-points in the IBZW, respectively) indicate that
the AHC, SHC, ANC and SNC obtained using $n_d$ = 50 converge to within a few \%. Indeed, the curves of AHC,
SHC, ANC and SNC as a function of energy (see Figs.  3-5 below) and also the curves of ANC and SNC as a
function of temperature (see Fig. 6 below) obtained with $n_d$ = 40 and 50 are indistinguishable. 
Moreover, the calculated AHC, SHC, ANC and SNC versus the inverse of the number ($N_k$) of $k$-points 
in the IBZW are plotted and fitted to a straight line to get the converged theoretical values listed 
in Table II below (i.e., the extrapolated values at $N_k = \infty$) (see Refs. [\onlinecite{Fuh11,Tung12}]) 
Note that the differences between the converged theoretical AHC, SHC, ANC and SNC values and the corresponding 
$n_d = 50$ values are within a few \%. 
As mentioned before, we also calculate the AHC and ANC of ferromagnetic bcc Fe and
the SNC of nonmagnetic fcc Pt for comparison. In the calculation of the AHC and ANC of bcc Fe,
we also adopt a very fine mesh of 360396 $k$-points on the magnetic IBZW (1/16 BZ). 
In the SHC and SNC calculations for fcc Pt, a very find grid of 253044 $k$-points on the magnetic IBZW
(1/16 BZ) is used. 

\section{Results and discussion}
The energetics of many possible magnetic configurations for Mn$_{3}$Sn in the hexagonal DO$_{19}$ 
structure has already been investigated with the {\it ab initio} density functional 
calculations \cite{Stic89,Sand96,Kueb14}. Therefore, in this paper we consider only two low-energy 
noncollinear triangular antiferromagnetic configurations for Mn$_{3}X$ ($X$ = Sn, Ge, Ga) 
[see Fig. 1(c) and Fig. 1(d) in Ref. \cite{Sand96}], namely, type A 
and type B configurations as illustrated in Fig. 1(c) and Fig. 1(d), respectively.
For comparison, the ferromagnetic state (FM) of Mn$_{3}$Ga with magnetic moments 
in the $\hat{x}$-direction is also investigated.

\begin{table}
\caption{Calculated total energy ($E_{t}$) and total spin magnetic moment ($m^s_{t}$) 
as well as averaged Mn spin magnetic moment ($m^s_{Mn}$)
for the A and B magnetic structures of Mn$_{3}X$ ($X$ = Sn, Ge, Ga). Total magnetic moments 
are parallel to the $\hat{x}$-axis in configuration A but to the $\hat{y}$-axis in configuration B. 
The $X$ atoms have a nearly zero magnetic moment (being less than 0.01 $\mu_{B}$) and thus are not listed.
Note that there are two formula units [i.e., 2(Mn$_{3}X$)] per unit cell.
For comparison, the results of the magnetic moment direction-constrained calculation for 
Mn$_3$Ga in configuration A (denoted A$^*$) and the ferromagnetic calculation for Mn$_3$Ga
with the magnetic moments in the $\hat{x}$-direction (denoted FM) are given as well.
Some previously reported total and Mn spin moments are also listed for comparison. }
 
\label{tab:spin}
\begin{ruledtabular}
\begin{tabular}{ccccc}
\multicolumn{1}{c}{}&\multicolumn{1}{c}{}&\multicolumn{1}{c}{$E_{t}$}&\multicolumn{1}{c}{$m^s_{\textrm{Mn}}$ }&\multicolumn{1}{c}{$m^s_{\textrm{t}}$ } \\

\multicolumn{1}{c}{}&\multicolumn{1}{c}{}&\multicolumn{1}{c}{(meV/cell)} &\multicolumn{1}{c}{($\mu_{B}/$atom)}&\multicolumn{1}{c}{($10^{-3}\mu_{B}/$cell)} \\
        \hline
Mn$_{3}$Sn & A & 0.0 &  3.13, 3.0\footnotemark[1]  &  0.1, 12\footnotemark[2] \\
           & B & 0.03 &  3.13  & 22 \\
Mn$_{3}$Ge & A &  0.0  &  2.70, 2.4\footnotemark[3]  &  0.9, 42\footnotemark[4] \\
           & B &  0.03 &  2.68  & 2.3, 30\footnotemark[5] \\
Mn$_{3}$Ga & A &  0.0  &  2.75, 2.4\footnotemark[6]  &  11.3\\
           & A$^*$& 0.00 & 2.75 & 0.6\\ 
           & B &  0.01 &  2.73  & 10.9\\ 
           & FM&  855 &  2.18  & 13019
\end{tabular}
\end{ruledtabular}
\footnotemark[1]{Ref.~\onlinecite{Zim71} (experiment),}
\footnotemark[2]{Ref.~\onlinecite{Nak15} (experiment),}
\footnotemark[3]{Ref.~\onlinecite{Kad71} (experiment),}
\footnotemark[4]{Ref.~\onlinecite{Kiy16} (experiment),}
\footnotemark[5]{Ref.~\onlinecite{Nay16} (experiment),}
\footnotemark[6]{Ref.~\onlinecite{Kre70} (experiment).}
\end{table}

\subsection{Magnetic properties}
The calculated total energies and spin magnetic moments are listed in
Table~\ref{tab:spin}, together with the reported experimental values.
Table~\ref{tab:spin} shows that in all three alloys, magnetic structure A has a lower energy than
magnetic structure B, although the total energy difference is in the order of $\sim$0.01 meV.
This agrees with the magnetic structure observed in earlier neutron diffraction
experiments on Mn$_3$Sn\cite{Tom83}, Mn$_3$Ge\cite{Tom83,Nag82} and  Mn$_3$Ga.\cite{Kre70}.
In an earlier DFT calculation for Mn$_3$Sn\cite{Sand96}, configuration A was also found to
be slightly lower in total energy than configuration B.
Nevertheless, the total energy difference is very small
and such a small energy difference is perhaps within the numerical uncertainty.
This small energy difference between the two configurations
is consistent with the experimental fact that the magnetic moments can be easily rotated
in the hexagonal plane by a small magnetic field\cite{Tom83,Nag82,Nak15,Nay16,Kiy16}.
In contrast, the total energy of the FM structure of Mn$_3$Ga is well above that of the two noncollinear
antiferromagnetic structures (Table ~\ref{tab:spin}).

The calculated Mn spin magnetic moments in all three Mn$_{3}X$ ($X$ = Sn, Ge, Ga) alloys are large, 
being $\sim3.0$ $\mu_{B}$, while the calculated spin magnetic moments of the $X$ atoms are nearly zero, 
being less than $\sim0.01$ $\mu_{B}$. Table~\ref{tab:spin} indicates that the 
calculated Mn spin magnetic moments agree fairly well with previous experiments.~\cite{Zim71,Nak15,Kre70}
Due to rather strong exchange coupling between large spin magnetic momonets on
the Mn moments, the N\'{e}el temperatures in these Mn-based alloys are as high
as 420 K in Mn$_{3}$Sn~\cite{Tom83}, 365 K in Mn$_{3}$Ge~\cite{Nag82} and 470 K in Mn$_{3}$Ga~\cite{Kre70}.

\begin{table*}
\caption{Calculated density of states at the Fermi level [$N(E_F)$] (states/eV/spin/f.u.), anomalous Hall conductivity
(AHC; $\sigma_{H}^A$) ($\sigma_{yz}^A$, $\sigma_{zx}^A$) and anomalous Nernst conductivity (ANC; $\alpha_{N}^A$) ($\alpha_{yz}^A$, 
$\alpha_{zx}^A$) as well as spin Hall conductivity (SHC) ($\sigma_{xy}^z$) and spin Nernst conductivity 
(SNC) ($\alpha_{xy}^z$) of  Mn$_{3}X$ ($X$ = Sn, Ge, Ga). For comparison, the calculated related 
properties of bcc Fe ($\sigma_{xy}^A$, $\alpha_{xy}^A$) and fcc Pt ($\sigma_{xy}^z$, $\alpha_{xy}^z$) 
are also listed.  Note that ANC and SNC listed here were calculated at temperature $T = 300$ K. The ANC
for Mn$_{3}$Sn in brackets were calculated at $T = 210$ K and the SNC for fcc Pt in brackets were calculated
at $T = 255$ K. For comparison, the results of the magnetic moment direction-constrained calculation for
Mn$_3$Ga in configuration A (denoted A$^*$) and the ferromagnetic calculation for Mn$_3$Ga
with the magnetic moments in the $\hat{x}$-direction (denoted FM) are given as well.
Some previous experimental results are also listed for comparison.}
\label{tab:ahc}
\begin{ruledtabular}
\begin{tabular}{ccccccccc}
& & $N(E_F)$ & $\sigma_{H}^A$ & $\sigma_{H}^A(\mu)'$ &$\alpha_{N}^A$ & $\sigma_{xy}^z$ & $\sigma_{xy}^z(\mu)'$ & $\alpha_{xy}^z$ \\
 & &                       &(S/cm)& (S/cm-eV) & (A/m-K) & ($\hbar$/e)(S/cm) & ($\hbar$/e)(S/cm-eV) & ($\hbar$/e)(A/m-K)\\
        \hline
Mn$_{3}$Sn & A &1.96 &-132, -68\footnotemark[1],-90\footnotemark[2] & -456 & -0.54 (-0.14),-0.39\footnotemark[2],-0.28\footnotemark[3] & 72 & -845 & 0.91\\
           & B &1.96 &-132, -126\footnotemark[1],-80\footnotemark[2]& -444 & -0.55 (-0.14),-0.32\footnotemark[2] & 74 & -834 & 0.88\\
Mn$_{3}$Ge & A &2.37 &-298, 310\footnotemark[4],150\footnotemark[5] &-9020 & -0.89 & 56 &  691 & 0.14\\
           & B &2.38 &-298, 380\footnotemark[4],500\footnotemark[5] &-8289 & -0.89 & 63 & 1000 & 0.09\\
Mn$_{3}$Ga & A &5.99 &-104                      & -3722 &2.41 &-219&-5323 & 1.01\\
           & A$^*$ &5.99 &-106                  & -3697 &2.38 &-219&-5134 & 1.17\\
           & B &5.99 &-103                      & -3953 &2.34 &-241&-3561 & 0.91\\
           & FM&6.82 & 181                      & -12836&-1.94&-678&-10601 & 0.44\\
 bcc Fe    &   &1.11 & 708,1200\footnotemark[5]  &-230  &0.50,1.8\footnotemark[6] & -  &  -   &  -  \\
 fcc Pt    &   &1.75 & - & - & - &2139 & 1214 & -1.09 (-0.91),-1.57\footnotemark[7]
\end{tabular}
\end{ruledtabular}
\footnotemark[1]{Ref.~\onlinecite{Nak15} (experiment at 50 K),}
\footnotemark[2]{Ref.~\onlinecite{Li16} (experiment at 210 K),}
\footnotemark[3]{Ref.~\onlinecite{Ikhl17} (experiment at 200 K),}
\footnotemark[4]{Ref.~\onlinecite{Kiy16} (experiment at 10 K),}
\footnotemark[5]{Ref.~\onlinecite{Nay16} (experiment at 2 K).}
\footnotemark[6]{Ref.~\onlinecite{Li16} (experiment at 300 K).}
\footnotemark[7]{Extracted from the experiment at 255 K [~\onlinecite{Mey17}].}
\end{table*}

Interestingly, in Mn$_{3}Y$ ($Y$ = Ir, Rh, Pt), the calculated total spin magnetic moment is zero 
and the two coplanar noncollinear T1 and T2 antiferromagnetic structures have the same total 
energy in the absence of the SOC.\cite{Chen14,Feng15} This suggests that the small total 
magnetic moment obtained with the SOC included in Mn$_{3}Y$ ($Y$ = Ir, Rh, Pt), is induced 
by the spin-canting caused by the Dzyaloshinskii-Moriya interaction (DMI) (i.e., the SOC).
In contrast, the nonzero total spin magnetic moment already exists 
in  Mn$_{3}X$ ($X$ = Sn, Ge, Ga) even without the SOC. For example, the total spin magnetic moment 
calculated without the SOC, is 4$\times10^{-3}$ $\mu_B$/cell along the $\hat{x}$-axis in magnetic structure 
A of Mn$_{3}$Sn, being larger than that in the presence of the SOC (Table~\ref{tab:spin}). 
This is because both magnetic structures A 
and B are orthorhombic with only one mirror plane (Fig. 1). In the A magnetic structure,
the mirror plane  $\mathcal{M}_x$ is parallel to the $yz$ plane [see Fig. 1(c)]. 
Since the total magnetic moment ${\bf m}_t$ is a pseudovector, $m_{t,y}$ and $m_{t,z}$ that are parallel
to the $yz$ plane transform, respectively, to $-m_{t,y}$ and $-m_{t,z}$ under the $\mathcal{M}_x$ reflection,
while $m_{t,x}$ remains unchanged. Consequently, $m_{t,y}$ and $m_{t,z}$ must be zero and only $m_{t,x}$ 
can be nonzero. In the B structure, the mirror plane  $\mathcal{M}_y$ is parallel to 
the $zx$ plane [see Fig. 1(d)], and a mirror reflection plus a translation ${\bf \tau} = (0,0,c/2)$ would
bring the magnetic structure back onto itself. In this case, only $m_{t,y}$ can be nonzero.
The calculated magnetic moments are consistent with these symmetry requirements (Table~\ref{tab:spin}).
In contrast, the T1 and T2 magnetic structures of Mn$_{3}Y$ are hexagonal and have three mirror 
planes,\cite{Chen14,Feng15} and thus all three components of the magnetic moments must be zero.
Furthermore, the calculated total energies of the T1 and T2 structures are the same.

\subsection{Anomalous Nernst effect}

Table~\ref{tab:ahc} lists the calculated anomalous Nernst conductivity ($\alpha_{ij}^A$), 
anomalous Hall conductivity ($\sigma_{ij}^A$) and density of states at the Fermi level
[$N(E_F)$] of Mn$_{3}X$ ($X$ = Sn, Ge, Ga) alloys. As discussed before, the AHC and ANC
are pseudovectors, just like the total magnetic moment. Thus, in the A (B) magnetic structure,
only $\alpha_{yz}^A$ ($\alpha_{zx}^A$) and $\sigma_{yz}^A$ ($\sigma_{zx}^A$) can be nonzero. 
This can also be seen from the ${\bf k}$-space distribution of the Berry curvature 
${{\bf\Omega}({\bf k})} = $[${\Omega^x({\bf k})},{\Omega^y({\bf k})},{\Omega^z({\bf k})}$], 
as displayed in Fig. 2 for configuration A of Mn$_3$Sn. Figure 2(a) shows that in the $k_xk_y$ (i.e., $k_z = 0$)
plane, ${\Omega^y({\bf k})}$ is an odd function of $k_x$ while ${\Omega^x({\bf k})}$ 
is an even function of $k_x$. In Fig. 2(b), ${\Omega^z({\bf k})}$ is found to be an odd function of $k_x$ 
while ${\Omega^x({\bf k})}$ is again an even function of $k_x$ in the $k_xk_z$ (i.e., $k_y = 0$) plane. 
Consequently, Eqs. (1) and (2) would indicate that $\sigma_{zx}^A$ and $\sigma_{xy}^A$ as well as
 $\alpha_{zx}^A$ and $\alpha_{xy}^A$ should be zero.
The present results (Table~\ref{tab:ahc}) are consistent with these symmetry properties.
It is also clear from Table~\ref{tab:ahc} that the AHC, ANC and $N(E_F)$ for both A and B configurations
are very similar. This is consistent with the fact that the two configurations have
nearly degenerate total energies and very similar magnetic properties (Table ~\ref{tab:spin}).

\begin{figure}
\includegraphics[width=8cm]{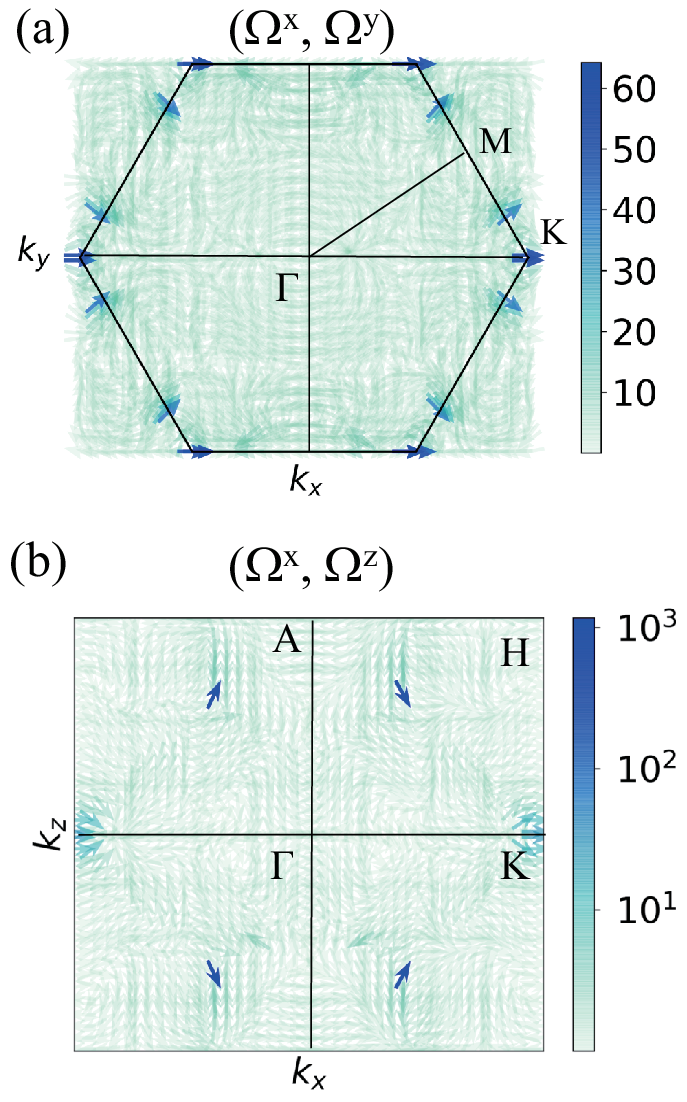}
\caption{\label{Omega} (Color online) Berry curvature 
[${\Omega^x({\bf k})},{\Omega^y({\bf k})},{\Omega^z({\bf k})}$] (in units of \AA$^{2}$) of 
configuration A Mn$_3$Sn. (a) ($\Omega^x, \Omega^y$) on the $k_xk_y$ ($k_z=0$) plane and 
(b) ($\Omega^x, \Omega^z$) on the $k_xk_z$ ($k_y=0$) plane. }
\end{figure}

The calculated $\alpha_{xy}^A$ and $\sigma_{xy}^A$ 
of iron metal are also listed there for comparison. Table ~\ref{tab:ahc} shows that 
the AHC of all the Mn$_{3}X$ alloys is rather large, being in the same order of magnitude as
that of ferromagnetic iron with a large net magnetic moment of 2.27 $\mu_B$/atom.
Remarkably, all the Mn$_{3}X$ alloys have a large ANC, which is up to 5 times
larger than that of Fe (Table ~\ref{tab:ahc}). This strongly suggests that these noncollinear
antiferromagnets would find promising applications in thermoelectric devices, heat nanosensors and also
spin caloritronics.

One may wonder whether the nonzero ANC and AHC are caused by the presence of the small net magnetization
in these noncollinear antiferromagnetic structures, as in the case of ferromagnets where the ANC and AHC
are proportional to the net magnetization. To address this issue, we perform the magnetic moment 
direction-constrained GGA calculation for the A structure of Mn$_{3}$Ga in order to make the total
magnetic moment vanished. The results of this calculation are listed in Tables I and II.
Table II shows that the resultant ANC and AHC remain nearly unchanged, although the net magnetic moment
is reduced by a factor of $\sim$20 (Table I). Moreover, we also carry out the GGA calculation for
Mn$_{3}$Ga in the ferromagnetic state (FM) with magnetization along the $\hat{x}$-axis. Interestingly,
although the total magnetic moment of the FM state is three orders of magnitude larger than
that of the A structure (Table I), the ANE gets reduced by 20 \%, compared with that of the A structure.  

\begin{figure}
\includegraphics[width=8cm]{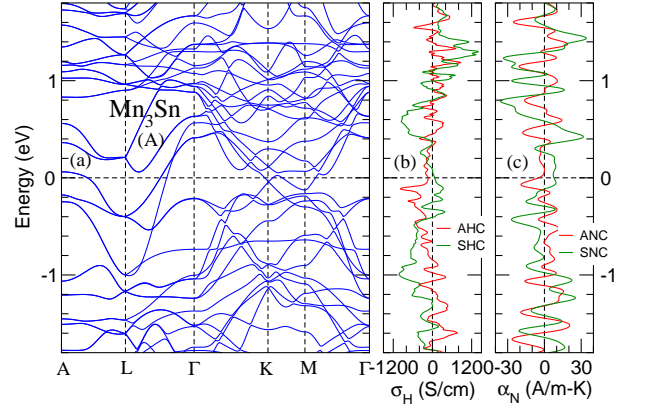}
\caption{\label{BSSn} (Color online) Mn$_3$Sn. (a) Relativistic band structure in magnetic structure A.
(b) Anomalous Hall conductivity (AHC) ($\sigma_{yz}^A$) and spin Hall conductivity (SHC) ($\sigma_{xy}^z$)
as well as (c) anomalous Nernst conductivity (ANC) ($\alpha_{yz}^A$) and spin Nernst conductivity (SNC) 
($\alpha_{xy}^z$) as a function of energy. The Nernst conductivities were calculated at $T = 300$ K.
The Fermi level is at the zero energy. Note that in (b) [(c)], the unit for the SHC [SNC] should be
($\hbar$/e)S/cm [($\hbar$/e)A/m-K].}
\end{figure}

\begin{figure}
\includegraphics[width=8cm]{Mn3XFig4.eps}
\caption{\label{BSGe} (Color online) Mn$_3$Ge. (a) Relativistic band structure in the A magnetic structure.
(b) Anomalous Hall conductivity (AHC) ($\sigma_{zx}^A$) and spin Hall conductivity (SHC) ($\sigma_{xy}^z$)
as well as (c) anomalous Nernst conductivity (ANC) ($\alpha_{zx}^A$) and spin Nernst conductivity (SNC) ($\alpha_{xy}^z$)
as a function of energy. The Nernst conductivities were calculated at $T = 300$ K.
The Fermi level is at the zero energy. Note that in (b) [(c)], the unit for the SHC [SNC] should be
($\hbar$/e)S/cm [($\hbar$/e)A/m-K].}
\end{figure}

\begin{figure}
\includegraphics[width=8cm]{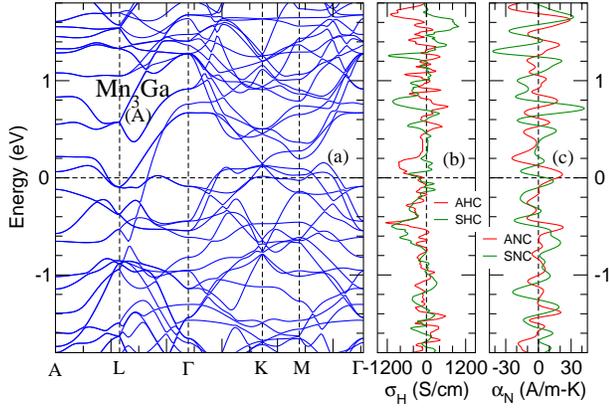}
\caption{\label{BSGa} (Color online) Mn$_3$Ga. (a) Relativistic band structure in the A magnetic structure.
(b) Anomalous Hall conductivity (AHC) ($\sigma_{zx}^A$) and spin Hall conductivity (SHC) ($\sigma_{xy}^z$)
as well as (c) anomalous Nernst conductivity (ANC) ($\alpha_{zx}^A$) and spin Nernst conductivity (SNC) 
($\alpha_{xy}^z$) as a function of energy. The Nernst conductivities were calculated at $T = 300$ K.
The Fermi level is at the zero energy. Note that in (b) [(c)], the unit for the SHC [SNC] should be
($\hbar$/e)S/cm [($\hbar$/e)A/m-K].}
\end{figure}

The calculated ANC ($\alpha_{yz}^A$) and AHC ($\sigma_{yz}^A$) of magnetic structure A 
as a function of the Fermi energy ($E_F$) as well as the relativistic band structure
are plotted in Fig. 3 for Mn$_3$Sn, in Fig. 4 for Mn$_3$Ge and in Fig. 5 for Mn$_3$Ga.
Figure 3 shows that for up to 0.33 eV above the $E_F$, the $\sigma_{yz}^A$ of Mn$_3$Sn is negative 
and rather flat with small ripples. However, if the Fermi energy is lowered to -0.114 eV, 
one sees a very pronounced negative peak in $\sigma_{yz}^A$. The peak $\sigma_{yz}^A$ 
value is -979 S/cm. To reach this energy level, the number of valence electrons
should be reduced by 0.206 per formula unit (f.u.), indicating substitution of
$\sim$20 \% of Sn by In or Ga. Examination of the calculated 
band-resolved Berry curvatures suggests that this peak arises predominantly
from the large $\Omega_{yz}$ on the top valence band in the vicinity of 
the gap at M-point [see Fig. 3(a)]. The shape of the $\sigma_{yz}^A$ versus $E_F$
curve in Mn$_3$Ge [Fig. 4(b)] is similar to that of Mn$_3$Sn [Fig. 3(b)], 
and this is understandable because both alloys are isoelectronic.
Mn$_3$Ga has roughly the same $\sigma_{yz}^A$ versus $E_F$ curve [Fig. 5(b)] as 
that of Mn$_3$Ge and Mn$_3$Sn except that the Fermi level is now about
0.25 eV lower mainly because Mn$_3$Ga has one less valence electron. 

To understand the features in the $\alpha_{yz}^A$ versus $E_F$ curve,
one should note that at low temperatures, Eq. (2) can be simplified as the Mott relation,
\begin{equation}
\alpha_{xy}^A = -\frac{\pi^2}{3}\frac{k_B^2T}{e}\sigma_{xy}^A(\mu)',
\end{equation}
which relates the ANC to the energy derivative of the AHC.
This Mott relation roughly explains why in Mn$_3$Sn [Fig. 3(c)] there is a
prominant peak in $\alpha_{yz}^A$ at -0.070 eV, where $\sigma_{yz}^A$
has a steep slope [Fig. 3(b)]. The peak $\alpha_{yz}^A$ value is as large as -2.24 A/m-K at 300 K. 
One could reach this point by reducing the valence electrons by 0.13 electron per Mn$_3$Sn,
i.e., by merely substituting 13 \% Sn with In or Ga. As mentioned before, $\sigma_{yz}^A$
is rather flat above the Fermi level [Fig. 3(b)], and this explains why
$\alpha_{yz}^A$ becomes nearly zero slightly above the $E_F$ [Fig. 3(c)].

We have also calculated the ANC of all the alloys as a function of temperature ($T$)
and the results are displayed in Fig. 6(a) together with the calculated $T$-dependent 
$\alpha_{xy}^A$ of bcc Fe. Figure 6(a) shows that at high temperature (300$\sim$400 K)
Mn$_{3}$Ga has a very large $\alpha_{yz}^A$, being up to $\sim$2.65 A/m-K which is 5 times
larger than that of bcc Fe. The $\alpha_{yz}^A$
of Mn$_{3}$Ga decreases steadily with decreasing $T$ and eventually approaches zero 
at $\sim$50 K. The magnitude of the ANC of Mn$_{3}$Sn and Mn$_{3}$Ge
is also large at high temperatures (e.g., $\sim$1.5 A/m-K at $T = 400$ K) but the sign
of the ANC is negative, being opposite to that of Mn$_{3}$Ga.
The magnitudes of the ANC of Mn$_{3}$Sn and Mn$_{3}$Ge decrease monotonically
as $T$ decreases and change sign at 175 K and 200 K, respectively. 
As $T$ further cools, the ANC of Mn$_{3}$Ge increases steadily and 
reaches to 0.72 A/m-K at 50 K, while that of Mn$_{3}$Sn stays around 
0.07 A/m-K with small fluctuations.
Because of their large ANC at room temperature [being at least one order of magnitude 
larger than that of bcc Fe (see Table II)], all three Mn$_{3}X$ alloys could serve 
as a thermoelectric material for spin caloritronics.

To examine the validity of the Mott relation [Eq. (5)], we calculated the
energy derivative of the AHC for all the alloys and bcc Fe, as listed in Table II.
The ANC at 100 K calculated using Eq. (5) and the energy derivatives
of the AHC are shown in Fig. 6(a). Figure 6(a) indicates that the ANC values
calculated this way agree in sign with those calculated directly using Eq. (2)
for Mn$_3$Sn, Mn$_3$Ge and bcc Fe. However, the magnitudes differ significantly.
At 300 K, the ANC for all Mn$_3X$ alloys estimated using Eq. (5) 
would differ in sign from those from Eq. (2) (listed in Table II).
We note that in the magnetized Pt and Pd, at 100 K the $\alpha_{xy}^A$ calculated
using the Mott relation [Eq. (5)] and directly from Eq. (2) agree quantitatively, 
and even at 300 K they agree with each other quite well.\cite{Guo14}

The band structures of magnetic structures A and B of all three alloys are
almost identical and thus their band structures for the B configuration
are not presented in this paper. Furthermore, the two magnetic configurations
for each alloy have similar AHC and ANC as a function of energy and hence
the AHC and ANC as a function of energy of the B configuration are not
displayed here either. The present band structures of Mn$_3$Sn (Fig. 3) and Mn$_3$Ge (Fig. 4) 
are in good agreement with the previous GGA results\cite{Zha17,Yan17}.
The present (Fig. 5) and previous\cite{Zha17} GGA band structures for Mn$_3$Ga
also agree quite well along all the high symmetry lines except the KM line where
the two band structures differ quite significantly.

As mentioned before, the AHE in Mn$_{3}$Sn and Mn$_{3}$Ge in noncollinear antiferromagnetic
states have been experimentally investigated by several groups.\cite{Nak15,Kiy16,Nay16,Li16} 
The calculated AHC (132 S/cm) for  Mn$_{3}$Sn in configuration B agrees well with the measured value 
(126 S/cm at 50 K) reported in Ref. [\onlinecite{Nak15}], although the theoretical AHC (132 S/cm)
in configuration A is nearly twice as large as the measured value (68 S/sm) (see Table II). 
The calculated AHC for  Mn$_{3}$Ge in both configuration A and B is also in good agreement with 
the experimental value at 10 K reported in Ref. [\onlinecite{Kiy16}], 
although for configuration B it is about 30 \% smaller than the measured one (500 S/cm at 2 K) 
presented in Ref. [\onlinecite{Nay16}] and for configuration A it is twice as large as 
the measured one\cite{Nay16}. All these suggest that the anomalous 
Hall effect in these alloys is dominated by the intrinsic mechanism due to 
the nonzero Berry curvatures in the momentum space.\cite{Naga10,Xiao10}
This is also the conclusion drawn in Ref. [\onlinecite{Li16}] based on the experimental
examination on the validity of the Wiedemann-Franz law. 
The AHC of Mn$_{3}$Sn (Mn$_{3}$Ge) presented in Table ~\ref{tab:ahc} is 
in excellent agreement with the GGA result of 133 (330) S/cm of 
Mn$_3$Sn (Mn$_{3}$Ge) reported in Refs. [\onlinecite{Zha17,Yan17}].   

However, unlike the AHE case, 
so far merely two papers very recently reported on the experiments on the ANE in Mn$_{3}$Sn.\cite{Li16,Ikhl17} 
It was found that the ANE signals are significant and easily detectable.\cite{Li16,Ikhl17}
Furthermore, the thermal and Nernst conductivities was found to correlate according to
the Wiedemann-Franz law, indicating the intrinsic origin of the ANE.
Overall, this is consistent with our finding of large intrinsic ANE in these alloys.
Also the measured and calculated ANC at $\sim$210 K agree in sign with respect to
that of AHC, although the measured ANC ($0.39$ and $0.28$ A/m-K) for configuration A
is about two times larger than the calculated ANC (0.14 A/m-K) (Table ~\ref{tab:ahc}). 
Nevertheless, experimentally, the ANC and AHC were found to decrease steadily
as  the $T$ is increased from 200 K to 400 K,\cite{Li16,Ikhl17} in contrast to the monotonical
increase of the calculated ANC with $T$ [Fig. 6(a)]. 
This significant discrepancy could arise from several reasons.
First of all, the temperature range of 200$\sim$ 400 K is close to the antiferromagnetic
transition ($T_N$) and consequently the magnetism gets weaker as the $T_N$ is approached. 
In the theoretical calculation, however, the $T =0$ magnetism is assumed and the $T$-dependence
enters only through the Fermi function [see Eq. (2)].
Secondly, although the ANC is calculated directly from the band structure [see Eq. (2)],
experimentally, the ANC cannot be measured directly and thus is estimated
using measurable longitudinal ($\rho_{ii}$) and Hall ($\rho_{ij}$) resistivities
as well as Seebeck ($S_{ii}$) and Seebeck-Nernst ($S_{ij}$) coefficients via\cite{Li16,Ikhl17}
\begin{equation}
\alpha_{yz}^A = -\frac{\rho_{zz}S_{yz}-\rho_{yz}S_{zz}}{\rho_{yy}\rho_{zz}-\rho_{yz}\rho_{zy}} 
\approx \frac{\rho_{zz}S_{yz}-\rho_{zz}S_{yz}}{\rho_{yy}\rho_{zz}}.
\end{equation}
Clearly, to obtain accurate estimated ANC, all these quantities must be accurately
measured on the same sample, but this often is not the case.
Given all these complications, we believe that the level of agreement between the experiment
and calculation is quite good. 
Table ~\ref{tab:ahc} shows that the experimental ANC of iron at 300 K also reported in [\onlinecite{Li16}] 
is $\sim$1.8 A/m-K, being four times larger than the present theoretical ANC (0.5 A/m-K).
Furthermore, a previous GGA calculation of the intrinsic ANC \cite{Weis13} of iron gave a value of 0.16 A/m-K, 
being more than two times smaller than the present GGA result. Further experiments on the ANE and
AHE on these alloys are clearly needed.

\begin{figure}[h]
\includegraphics[width=8cm]{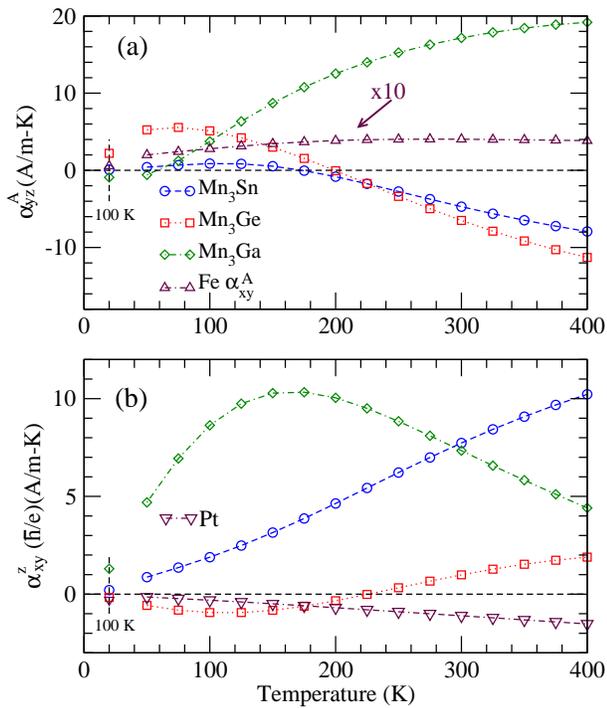}
\caption{\label{Tdep} (color online) (a) Anomalous Nernst conductivity (ANC) ($\alpha^A$)
and (b) spin Nernst conductivity (SNC) ($\alpha_{xy}^S$) as a function of temperature.
The solid symbols denote the values of the ANC and SNC at 100 K calculated
using the Mott relations [Eq. (5) and Eq. (7)] and the energy 
derivatives of AHC and SHC listed in Table II, respectively.}
\end{figure}

\subsection{Spin Nernst effect}

The SNC ($\alpha_{xy}^s$; $s,i,j = x,y,z$) and SHC ($\sigma_{xy}^s$) are third-order tensors.
A recent symmetry analysis\cite{Zha17} showed that only elements $\sigma_{yz}^x (\sigma_{zy}^x)$,
$\sigma_{xz}^y (\sigma_{zx}^y)$ and $\sigma_{xy}^z (\sigma_{yx}^z)$ can be nonzero. 
Furthermore, the {\it ab initio} calculations of the SHC of Mn$_{3}X$ ($X$ = Sn, Ge, Ga)\cite{Zha17}
indicated that only $\sigma_{xy}^z$ and $\sigma_{yx}^z$ are significantly nonzero. Therefore, 
in this paper we consider only $\alpha_{xy}^z$ and $\sigma_{xy}^z$.
The calculated $\alpha_{xy}^z$ and $\sigma_{xy}^z$ of all Mn$_{3}X$ alloys are listed in Table~\ref{tab:ahc}. 
The $\alpha_{xy}^z$ and $\sigma_{xy}^z$ of platinum metal\cite{Guo14} are also listed 
there for comparison. Table ~\ref{tab:ahc} shows that the SHC of the Mn$_{3}X$ alloys 
is rather small, compared to that of  platinum, which has the largest intrinsic SHC among 
transition metals.\cite{Guo08,Hoff13} Remarkably, the SNC of Mn$_{3}$Sn and Mn$_{3}$Ga 
is very large, being as large as that of Pt (Table ~\ref{tab:ahc}). 
Mn$_{3}$Ge also has a larger SNC than platinum. This shows that noncollinear antiferromagnets
Mn$_{3}X$ ($X$ = Sn, Ge, Ga) would be very useful materials for spin thermoelectric devices
and spin caloritronics, just like Pt metal for spintronics. 

The calculated SNC ($\alpha_{xy}^z$) and SHC ($\sigma_{xy}^z$) as a function of the
Fermi energy ($E_F$) of Mn$_3$Sn, Mn$_3$Ge and Mn$_3$Ga are displayed 
in Fig. 3, Fig. 4 and Fig. 5, respectively.
Figures 3(b) and 4(b) show that in both Mn$_3$Sn and Mn$_3$Ge the $\sigma_{xy}^z$ in 
the vicinity of the $E_F$ is rather small, thus resulting in a small value at the $E_F$
(Table ~\ref{tab:ahc}). Nevertheless, the $\sigma_{xy}^z$ in Mn$_3$Ge has a broad prominant 
peak near -0.30 eV, and the peak value is as large as -750 ($\hbar$/e)S/cm [Fig. 4(b)].
This peak can be reached by a reduction of the valence electrons of $\sim$1.0 e/f.u.
For Mn$_3$Ga which has one less valence electron, the $E_F$ is lowered to just below this peak [Fig. 5(b)],
thus resulting in a much larger $\sigma_{xy}^z$ value (Table ~\ref{tab:ahc}).

To understand the features in the $\alpha_{yz}^z$ versus $E_F$ curve,
one should note again that Eq. (4) would be reduced to the simple Mott relation at low temperatures,
\begin{equation}
\alpha_{xy}^z = -\frac{\pi^2}{3}\frac{k_B^2T}{e}\sigma_{xy}^z(\mu)',
\end{equation}
which relates the SNC to the energy derivative of the SHC.
This Mott relation roughly explains why in Mn$_3$Sn [Fig. 3(c)] the $\alpha_{yz}^z$ has a
broad plateau from -0.09 eV to 0.23 eV around the $E_F$, where $\sigma_{yz}^z$
has more or less a constant negative slope [Fig. 3(b)]. The plateau $\alpha_{yz}^z$ value 
is about 1.1 ($\hbar$/e)A/m-K at 300 K.
In Mn$_3$Ge, the $\alpha_{yz}^z$ is rather small in the vicinity of the $E_F$
because $\sigma_{yz}^z$ is rather flat (and small) (Fig. 4). 
Nevertheless, the $\alpha_{yz}^z$ has a prominant negative peak at -0.21 eV [Fig. 4(c)] 
where $\sigma_{yz}^z$ has a steep slope [Fig. 4(b)]. Within the rigid band model,
the $\alpha_{yz}^z$ peak could be reached by reducing the number of valence electrons
by $\sim$0.51 e/f.u. In Mn$_3$Ga, the $E_F$ sits on the upper side of the pronounced 
peak at -0.035 eV and thus $\alpha_{yz}^z$ is large, being as large as 1.9 ($\hbar$/e)A/m-K at 300 K. 
Again, this is because the $\sigma_{yz}^z$ has a steep slope at -0.035 eV [Fig. 5(b)].

Very recently, the SNE in platinum was studied experimentally and a large spin Nernst 
angle ($\theta_{SN}$) was observed.\cite{Mey17} The spin Nernst angle is comparable in size 
but opposite in sign to the spin Hall angle ($\theta_{SH}$) 
with $\theta_{SH}$/$\theta_{SN} = -0.5$ at 255 K.\cite{Mey17} 
It can be shown that $\alpha_{xy}^z = -\sigma_{xy}^z S_{yy}/(\theta_{SH}/\theta_{SN}$).
Using the theoretical $\sigma_{xy}^z = 2139$ ($\hbar$/e)(S/cm) (Table~\ref{tab:ahc}) 
and measured Seebeck coefficient $S_{yy} = -3.67$ $\mu$V/K at 255 K,\cite{Moor73} 
we would obtain an estimated experimental $\alpha_{xy}^z = -1.57$ ($\hbar$/e)(S/cm), 
agreeing quite well with the calculated 
value of -0.91 ($\hbar$/e)A/m-K (Table~\ref{tab:ahc}). 

In Fig. 6(b), the calculated $T$-dependence of the SNC for all three alloys as well 
as Pt metal are displayed. Fig. 6(b) shows that the magnitude of the SNC of Mn$_{3}$Sn
is very large at high temperatures (e.g., $\sim$1.2 ($\hbar$/e)A/m-K at $T = 400$ K). 
Nevertheless, the SNC decreases monotonically as the $T$ decreases down to 50 K.
Interestingly, the SNC of Pt has a very similar $T$-dependence, albeit with a 
much smaller magnitude and an opposite sign. In contrast, Mn$_{3}$Ga has a smaller SNC at high temperatures 
(e.g., $\sim$0.61 ($\hbar$/e)A/m-K at $T = 400$ K). However, the SNC of Mn$_{3}$Ga
increases steadily as the $T$ is lowered, and it reaches its maximum of $\sim$1.34
($\hbar$/e)A/m-K at $T = 175$ K. It then decreases monotonically as the $T$ further decreases.
Mn$_{3}$Ge has a small SNC at high temperatures (e.g., $\sim$0.26 
($\hbar$/e)A/m-K at $T = 400$ K). The SNC of Mn$_{3}$Ge decreases gradually 
as the $T$ decreases and changes sign at $T = 225$ K. After passing 225 K, it further decreases
as the $T$ is lowered to 125 K, and it then increases slightly as the $T$ decreases to 50 K.
 
We calculated the energy derivative of the SHC for all the alloys and fcc Pt, as listed 
in Table II, in order to examine the validity of the Mott relation [Eq. (7)].
The SNC at 100 K calculated using Eq. (7) and the energy derivatives
of the SHC are shown in Fig. 6(b). Figure 6(b) indicates that all the SNC values
calculated this way agree in sign with those calculated directly using Eq. (4).
For fcc Pt, the SNC values [-0.30 and -0.31 ($\hbar$/e)S/cm] agree rather well.
This level of agreement [-0.89 and -1.11 ($\hbar$/e)S/cm] is maintained even at 300 K.
For all three Mn compounds Mn$_{3}$X, the SNC values estimated using Mott relation [Eq. (7)]
agree rather well with that calculated directly using Eq. (4) [Fig. 6(b)].

\section{Conclusions}
We have studied theoretically the ANE, a phenomenon having the same origin as
the AHE, and also the SNE as well as the AHE and SHE
in noncollinear antiferromagnetic Mn$_3X$ ($X$ = Sn, Ge, Ga) 
based on {\it ab initio} relativistic band structure calculations.
As references, we also calculate the ANC and AHC of ferromagnetic iron 
as well as the SNC of platinum metal. Fascinatingly, the calculated ANC at room temperature (300 K)
for all three alloys is large, being up to 5 times larger than that of iron.
Further, the calculated SNC for Mn$_3$Sn and Mn$_3$Ga is also large, being as large as
that of platinum. This suggests that these antiferromagnets would
be useful materials for thermoelectronic devices and spin caloritronic devices.
The calculated ANC of Mn$_3$Sn and iron are in reasonably good agreement with
the very recent experiments\cite{Li16}. The calculated SNC of platinum also agree well with
the very recent experiments\cite{Mey17} in both sign and magnitude.
The calculated thermoelectric and thermomagnetic properties are analyzed
in terms of the band structures as well as the energy-dependent AHC, ANC, SNC and SHC
via the Mott relations. We hope that our interesting theoretical 
results would stimulate further experimental works on these noncollinear antiferromagnets.

\section*{Acknowledgments}
The authors acknowledge support from the Ministry of Science and Technology and the Academia Sinica
of The R.O.C. as well as  
the NCTS and the Kenda Foundation in Taiwan. G.Y.G thanks Qian Niu for stimulating discussions.

\end{document}